\documentclass[12pt]{iopart}
  \expandafter\let\csname equation*\endcsname\relax
  \expandafter\let\csname endequation*\endcsname\relax
 \usepackage{amsmath}
\usepackage[matrix,frame,arrow]{xypic}
\usepackage[pdfstartview=FitH]{hyperref}
\usepackage{pifont}
\hypersetup{
    colorlinks=true,       	% false: boxed links; true: colored links
    linkcolor=red,          	% color of internal links
   citecolor=magenta,        % color of links to bibliography
    filecolor=magenta,      	% color of file links
    urlcolor=cyan,           	% color of external links
    runcolor=cyan
}
\usepackage[pdftex]{color}
\usepackage{braket}%Dirac Notation in QM
\usepackage{enumerate}
\usepackage[normalem]{ulem}
\usepackage{subfigure}
\usepackage{overpic}

\usepackage{multirow}

\newcommand{\be}{\begin{equation}}
\newcommand{\ee}{\end{equation}}
\newcommand{\ba}{\begin{eqnarray}}
\newcommand{\ea}{\end{eqnarray}}
\newcommand{\prj}[1]{|#1\rangle\langle #1|}
\newcommand{\ignore}[1]{}

\begin{document}
\bibliographystyle{plain}
\title[]{Robust control of long distance entanglement in disordered spin chains}

\author{Jian Cui$^{1,2}$ and Florian Mintert$^{1,2}$}

\address{$^1$QOLS, Blackett Laboratory, Imperial College London, SW7 2BW, United Kingdom}
\address{$^2$Freiburg Institute for Advanced Studies, Albert Ludwigs University of Freiburg, Albertstra{\ss}e 19, 79104 Freiburg, Germany}

\begin{abstract}
We derive temporally shaped control pulses for the creation of long-distance entanglement in disordered spin chains.
Our approach is based on a time-dependent target functional and a time-local control strategy
that permits to ensure that the description of the chain in terms of matrix product states is always valid.
With this approach, we demonstrate that long-distance entanglement can be created even for
substantially disordered interaction landscapes.

\end{abstract}

\vspace{2pc}

\maketitle

\section{Introduction}

Many elementary tasks of quantum information processing can be performed on small scales with existing technology.
For example, state tomography \cite{tomography} on a single qubit is routinely done in many laboratories,
and the number $15$ has been factorized with a quantum device \cite{shor:nmr}.
Realizing these tasks on larger scale is one of the most pressing challenges in nowadays research on engineered quantum systems:
characterizing the state of many qubits is an actively pursued problem even on the theoretical side \cite{PhysRevLett.111.020401},
and factorizing $77$ or $187$ is still impossible with our available technology.
Similarly, entangled states of two qubits can be prepared with many systems \cite{nature_86_497,Nature_504_461}, but most setups are not scalable;
the number of entangled photons is limited by the increasingly low probabilities of spontaneous events \cite{nat_photonics_6_225},
and satisfactory scaling has so far been demonstrated for trapped ions only \cite{14enttrappedions}.

A central difference between trapped ions and other systems is that ions interact via long-range interactions, what facilitates the creation of strongly entangled states.
Most other systems, however, are limited by rapidly decreasing interactions, and this disadvantage easily compensates the added value of the long coherence times \cite{NVtime} that can be found {\it e.g.} in impurities of solid state lattices like nitrogen-vacancy ($NV$) centres.
These unfavourable interaction properties can be improved if auxiliary quantum systems are available to mediate interactions \cite{Meijer:2006fk,PhysRevLett.102.070501,doi:10.1021/nl102066q},
and the establishment of long-distance entanglement via chains of auxiliary spins seems to be a very promising route \cite{Bukach:2010aa,PhysRevLett.107.150503}.

Most approaches, however, rely on perfectly ordered chains \cite{PhysRevA.81.022321, PhysRevLett.87.017901,Murphy}, whereas any implantation of auxiliary spins is likely to result in a slightly disordered chain with non-uniform interactions.
Our goal is to devise temporally shaped control fields that permit to establish long-distance entanglement independently of the specific realization of such disorder.

\section{Numerical tool: Matrix Product States.}
Compensation of disorder through suitably designed control fields is well established and has been demonstrated abundantly.
In particular numerical pulse design \cite{krotov1995global,Khaneja2005296} has proven very successful.
In our current goal, however, numerical approaches suffer from the inherent growth of complexity of composite quantum systems with the number of constituents.
We therefore resort to a description in terms of matrix-product-states (MPS);
that is, a state $\ket{\Psi}$ of $N$ spins is parametrised in terms of matrices $A_{i_j}$ \cite{Schollwoeck201196,RevModPhys.77.259} via the prescription
\begin{equation}
\ket{\Psi}=\sum_{i_1\cdots i_N}\tr\left[A_{i_1}\cdots A_{i_N}\right]\ket{i_1\cdots i_N}\ .
\end{equation}
The dimensions of these matrices (bond dimension) limit the overall entanglement of states that can be described with this ansatz, but a bond dimension between $20$ and $30$ is enough for the present purposes.
MPS permit to treat systems of several hundreds of spins \cite{Perez-Garcia:2007:MPS:,doi:10.1080/14789940801912366}. Based on the MPS description and the underlying variational ansatz \cite{Schollwoeck201196}, advanced numerical algorithms for the simulation of large systems with weakly entangled states have been developed \cite{PhysRevLett.93.040502,PhysRevLett.107.070601}, and such efficient descriptions provide an extremely promising starting point for the control of large quantum systems \cite{Tommaso}.
In general, MPS can simulate the time evolution of many-body system efficiently and accurately only for a short period of time due to the increase of entanglement between any two blocks of components \cite{Entgrowth,Entgrowth_PRX}.
Despite their limitation to describe weakly entangled states, however, MPS are a viable option for our purpose since we target the creation of strong entanglement among few distant spins.
Whereas MPS fail to describe strongly entangled states of $N$ spins without loosing their favourable scaling in $N$,
they are perfectly capable to describe the state of an $N$ spin system in which only a subset $M\ll N$ is strongly entangled.
We will therefore strive for a control strategy that ensures that the $N$-spin system is weakly entangled during the entire time-window while $M$-body entanglement is being enhanced.

\section{Control Scheme}
\label{sec:control scheme}

Despite the favourable scaling of the computational effort with $N$, simulating a system with $N\gg 1$ spins in terms of MPS is a numerically expensive endeavour.
Typical pulse shaping algorithms, however, rely on an iterative refinement that requires many repeated propagations \cite{krotov1995global,Khaneja2005296}, which pushes the problem from hard to practically impossible.
In order to be able to treat sufficiently large spin chains, we therefore resort to a variation of Lyapunov control \cite{PhysRevLett.105.020501} that permits to identify a good pulse with a single propagation only.
Normally Lyapunov control is based on the identification of the control field that maximizes the increment of the selected goal at each instance of time.
The present goal is the creation of entanglement, and since entanglement is independent of local spin orientations, we can always choose our target functional to be independent of single-spin dynamics.

Suppose initially the system is in a completely separable state and there is no direct interaction between the end spins;
in this case the present time-local control scheme will not identify any control Hamiltonian that results in an increase of 
pairwise entanglement between site
$1$ and $N$ (defined more rigorously later in Eq. \ref{eq:tau});
only at a later stage when some entanglement has been built up,
will a suitable control Hamiltonian be identified.
We will therefore define a time-dependent target that is such that one can always find a suitable control Hamiltonian, and that will eventually coincide with the desired entanglement measure. 
If we consider a chain with nearest neighbour interaction only, then the only goal that is initially achievable is entanglement between neighbouring spins, say spin $1$ and $2$.
Once this goal is achieved, one may strive for the creation of entanglement between spin $1$ and $3$.
Such an entanglement swapping scheme can be realized with a sequence of $N-1$ time-intervals;
pairwise entanglement between site $1$ and $j$ is targeted 
in the $j-1^{st}$ interval,
and the end of this interval is reached once a satisfactory value for the target 
has been reached.

Such target functionals should favour pairwise entanglement
between two selected spins, and, additionally, penalize entanglement shared by any other spin in order to ensure validity of the description in terms of MPS.
The entanglement between spin $i$ and the rest of the system can be characterised in terms of the purity $S(\varrho_i)=1-\tr\varrho_i^2$ of the reduced density matrix $\varrho_i$.
A target functional that is maximised if spins $i$ and $j$ form a Bell state can be chosen as $S(\varrho_i)+S(\varrho_j)-\mu S(\varrho_{i,j})$.
The first two terms favour entanglement shared by spins $i$ and $j$;
the third term takes into account that mixed states tend to be weakly entangled or separable \cite{PhysRevB.71.153105}.
For $\mu=2$, this is a lower bound \cite{PhysRevA.75.052302} to the concurrence \cite{PhysRevLett.80.2245} of mixed states, which is particularly good for weakly mixed states \cite{PhysRevA.72.012336};
for use of control target, however, we found that lower values of $\mu$ are favourable, and we will use $\mu=1/5$ later-on.
With an additional penalty for entanglement shared by spins different than $i$ or $j$,
we arrive at the target
\be
\tau_{ij}=S(\varrho_i)+S(\varrho_j)-\mu S(\varrho_{i,j})-\sum_{k\neq i,j}\alpha_kS(\varrho_k)\ .
\label{eq:tau}
\ee
where the non-negative scalars $\alpha_k$ permit to choose the emphasis on the penalty for spin $k$. 

For the specific physical situations to be considered, we will limit ourselves to single-spin control Hamiltonians, as realistically available means of control like microwave or laser-fields induce such single-spin dynamics.
In regular Lyapunov control, the control Hamiltonian is constructed based on the time-derivative of the target functional,
but since the present functionals are invariant under single-spin dynamics,  $\dot\tau_{ij}$ does not permit to construct an optimal control Hamiltonian.
It is, however, possible to consider the curvature $\ddot\tau$ rather than the increase $\dot\tau$ of a target functional $\tau$ to read off an instantaneously optimal control Hamiltonian
\cite{PhysRevLett.105.020501,1367-2630-13-7-073001}.
The curvatures $\ddot\tau$ are defined in terms of the first two temporal derivatives
\ba
\dot\varrho_j&=&-i\ \tr_{\bar j}\big[H,\prj\Psi\big]\ \mbox{and}\label{rhodot}\\
\ddot\varrho_j&=&-i\ \tr_{\bar j}\big[\dot H,\prj\Psi\big]-\tr_{\bar j}\big[H,\big[H,\prj\Psi\big]\big]\ ,\label{rhoddot}
\ea
of the reduced density matrices $\varrho_j$,
which, in turn depend on the state $\ket{\Psi}$ of the full chain,
and the chain's Hamiltonian $H=H_s+H_c(t)$ \cite{PhysRevA.88.032306} comprised of the static system Hamiltonian $H_s$ and the to-be-designed time-dependent control Hamiltonian $H_c(t)$;
`$\tr_{\bar j}$' denotes the partial trace over all spins but spin $j$.
Since the tunable control Hamiltonian does not contain any interaction terms,
it can be written as
\begin{equation}
H_c(t)=\sum_{i=1}^{N}\sum_{\theta}g_i^{(\theta)}(t)\ \sigma_i^{(\theta)}\ ,
\label{eq:Hc}
\end{equation}
in terms of the usual Pauli matrices $\sigma_i^{(\theta)}$ for spin half systems where $\theta=\{x,y,z\}$, or different operators that correspond to implementable Hamiltonians.

Eqs. \eqref{rhodot} and \eqref{rhoddot} suggest that $\ddot\tau_{ij}$ depends bi-linearly on $H_c(t)$, and that it depends linearly on $\dot H_c(t)$.
Similarly to the reason why $\dot\tau_{ij}$ does not depend of $H_c(t)$, however, one may see that $\ddot\tau_{ij}$ depends on $H_c(t)$ only linearly, and that it does not depends on $\dot H_c(t)$ at all;
bilinear terms in $H_c(t)$ and linear terms in $\dot H_c(t)$ correspond to local unitary dynamics, {\it i.e.} dynamics that $\tau_{ij}$ is invariant under.
Given the linear dependence in $H_c(t)$, one can thus express $\ddot\tau_{ij}$ as
\be
\ddot\tau_{ij}=\sum_{p=1}^N\sum_{\theta}\frac{\partial \ddot\tau_{ij}(\Psi)}{\partial g_p^{(\theta)}}g_p^{(\theta)}
+\left.\ddot\tau_{ij}\right|_{g_p^{\theta}=0\ \forall\ \theta}\ .
\ee
The maximium of $\ddot\tau_{ij}$ under the constraint that the magnitude $\sum_{k}(g_i^{(\theta)})^2$ of a local control Hamiltonian is limited by some maximally admitted value is obtained for
\be
\left. g_i^{(\theta)}\right|_{opt}=Z_i\frac{\partial \ddot\tau_{1j}(\Psi)}{\partial g_i^{(\theta)}}\ ,
\label{eq:optimalparameters}
\ee
with the normalization constant $Z_i$ chosen such that the control does not exceed its admitted strength.
This optimal choice can be constructed as analytic function of the system state $\ket{\Psi}$ and the system Hamiltonian,
so that at any instance during the propagation the optimal choice of $H_c$ is available \cite{PhysRevLett.105.020501,PhysRevA.88.032306}.
For the actual propagation, it is practical not to work with time-dependent Hamiltonians, but rather choose the control parameters constant during some short time interval $\Delta$ and update this choice after each multiple of this period \cite{PhysRevLett.93.040502}.

Since $\tau$ is based on single-spin and two-spin reduces density matrices only, and the Hamiltonian contains only single-spin and two-spin terms, the optimal control Hamiltonian is characterised completely in terms of up to three-spin reduced density matrix only that can be constructed efficiently from the MPS. 
In fact, for $\mu=0$, $\tau_{ij}$ defined in Eq.~\eqref{eq:tau} is defined in terms of single-spin reduced density matrices only,
so that the optimal control Hamiltonian can be constructed exclusively in terms of two-spin reduced density matrices.

\section{The Ising chain}

To be specific, let us consider the example of a spin chain with an Ising interaction
\begin{equation}
H_{s}(\vec J)=\sum_{i=1}^{N-1} J_i \sigma_i^z \sigma_{i+1}^z\ ,
\label{eq:Hs}
\end{equation}
that is characterized by a vector $\vec J$ that contains the coupling constants for the $N-1$ nearest neighbour interactions.
For this specific Hamiltonian, the $\sigma_z$ components of the optimal control Hamiltonian vanish (since they commute with $H_{s}$), and the $\sigma_x$ and $\sigma_y$ components are obtained from 
\be\begin{array}{rcl}
\displaystyle\frac{\partial\ddot S(\rho_i)}{\partial g_i^{\theta}}&=& 
J_{i-1}(\rho_i^{\theta}\rho_{i-1,i}^{z,z}-\rho_i^{z}\rho_{i-1,i}^{z,\theta})+J_i(\rho_i^{\theta}\rho_{i,i+1}^{z,z}-\rho_i^{z}\rho_{i,i+1}^{\theta,z})\vspace{.3cm}\nonumber\\
\displaystyle\frac{\partial\ddot S(\rho_i)}{\partial g_{i+ 1}^{\theta}}&=&
J_{i}(\rho_i^{\phi}\rho_{i,i+1}^{\theta,\phi}-\rho_i^{\theta}\rho_{i,i+1}^{\phi,\phi})\ ,\vspace{.3cm}\nonumber\\
\displaystyle\frac{\partial\ddot S(\rho_i)}{\partial g_{i- 1}^{\theta}}&=&
J_{i- 1}(\rho_i^{\phi}\rho_{i- 1,i}^{\phi,\theta}-\rho_i^{\theta}\rho_{i- 1,i}^{\phi,\phi})\ ,\vspace{.3cm}\nonumber\\
\displaystyle\frac{\partial\ddot S(\rho_{12})}{\partial g_1^{\theta}}&=&0\ ,\vspace{.3cm}\nonumber\\
\displaystyle\frac{\partial\ddot S(\rho_{12})}{\partial g_2^{\theta}}&=&
\displaystyle J_{2} \sum_{\Delta=0}^3  (\rho_{1,2}^{\Delta, \theta}\rho_{1,2,3}^{\Delta, z,z}-\rho_{1,2}^{\Delta, z}\rho_{1,2,3}^{\Delta, \theta,z})\ ,\vspace{.3cm}\nonumber\\
\displaystyle\frac{\partial\ddot S(\rho_{12})}{\partial g_3^{\theta}}&=&
\displaystyle J_{2}\sum_{\Delta=0}^3(\rho_{1,2}^{\Delta, \phi}\rho_{1,2,3}^{\Delta, \theta,\phi}-\rho_{1,2}^{\Delta, \theta}\rho_{1,2,3}^{\Delta, \phi,\phi})\ , \nonumber
\end{array}\ee
and (for $j>2$) from
\be\begin{array}{rcl}
\displaystyle\frac{\partial\ddot S(\rho_{1j})}{\partial g_1^{\theta}}&=&
\displaystyle J_1\sum_{\Delta=0}^3(\rho_{1,j}^{\theta,\Delta}\rho_{1,2,j}^{z,z,\Delta}-\rho_{1,j}^{z,\delta}\rho_{1,2,j}^{\theta,z,\Delta})\ ,\vspace{.3cm} \nonumber\\
\displaystyle\frac{\partial\ddot S(\rho_{1j})}{\partial g_2^{\theta}}&=&
\displaystyle J_1\sum_{\Delta=0}^3(\rho_{1,j}^{\phi,\Delta}\rho_{1,2,j}^{\theta,\phi,\Delta}-\rho_{1,j}^{\theta,\Delta}\rho_{1,2,j}^{\phi,\phi,\Delta})\ ,\vspace{.3cm} \nonumber\\
\displaystyle\frac{\partial\ddot S(\rho_{1j})}{\partial g_{j+ 1}^{\theta}}&=&
\displaystyle J_{j}\sum_{\Delta=0}^3(\rho_{1,j}^{\Delta, \phi}\rho_{1,j,j+1}^{\Delta, \theta,\phi}-\rho_{1,j}^{\Delta, \theta}\rho_{1,j,j+1}^{\Delta, \phi,\phi}),\vspace{.3cm} \nonumber\\
\displaystyle\frac{\partial\ddot S(\rho_{1j})}{\partial g_{j- 1}^{\theta}}&=&
\displaystyle J_{j-1}\sum_{\Delta=0}^3(\rho_{1,j}^{\Delta, \phi}\rho_{1,j-1,j}^{\Delta, \phi,\theta}-\rho_{1,j}^{\Delta, \theta}\rho_{1,j-1,j}^{\Delta, \phi,\phi})\ ,\vspace{.3cm} \nonumber\\
\displaystyle\frac{\partial\ddot S(\rho_{1j})}{\partial g_j^{\theta}}&=&\displaystyle\sum_{\Delta=0}^3 J_{j-1} (\rho_{1,j}^{\Delta, \theta}\rho_{1,j-1,j}^{\Delta, z,z}-\rho_{1,j}^{\Delta, z}\rho_{1,j-1,j}^{\Delta, z,\theta}) + J_{j}(\rho_{1,j}^{\Delta, \theta}\rho_{1,j,j+1}^{\Delta, z,z}-\rho_{1,j}^{\Delta, z}\rho_{1,j,j+1}^{\Delta, \theta,z})\ , \nonumber
\end{array}\ee
where
$\{\theta,\phi\}=\{x,y\}$, and $\theta \neq \phi$.
$\rho_i^{\theta_i}=\tr(\sigma_i^{\theta_i}\rho)$,
$\rho_{i,j}^{\theta_i,\theta_j}=\tr(\sigma_i^{\theta_i}\sigma_j^{\theta_j}\rho)$  and
$\rho_{i,j,k}^{\theta_i,\theta_j,\theta_k}=\tr(\sigma_i^{\theta_i}\sigma_j^{\theta_j}\sigma_k^{\theta_k}\rho)$
are the expectation values of the spin operators
$\sigma_{\theta_i}$, $\sigma_{\theta_i}\otimes\sigma_{\theta_j}$ and $\sigma_{\theta_i}\otimes\sigma_{\theta_j}\otimes\sigma_{\theta_k}$ 
for spins $i,j$ and $k$.
In the case of spins at the end of the chain, {\it i.e.} $i=1$, $i=N$ or $j=N$, it is implied that $J_0=J_{N}=0$ since there is no corresponding interaction partner.

\subsection{Ideal Ising chain.}

\begin{figure}[h]
\begin{center}
\includegraphics[width=0.9\linewidth]{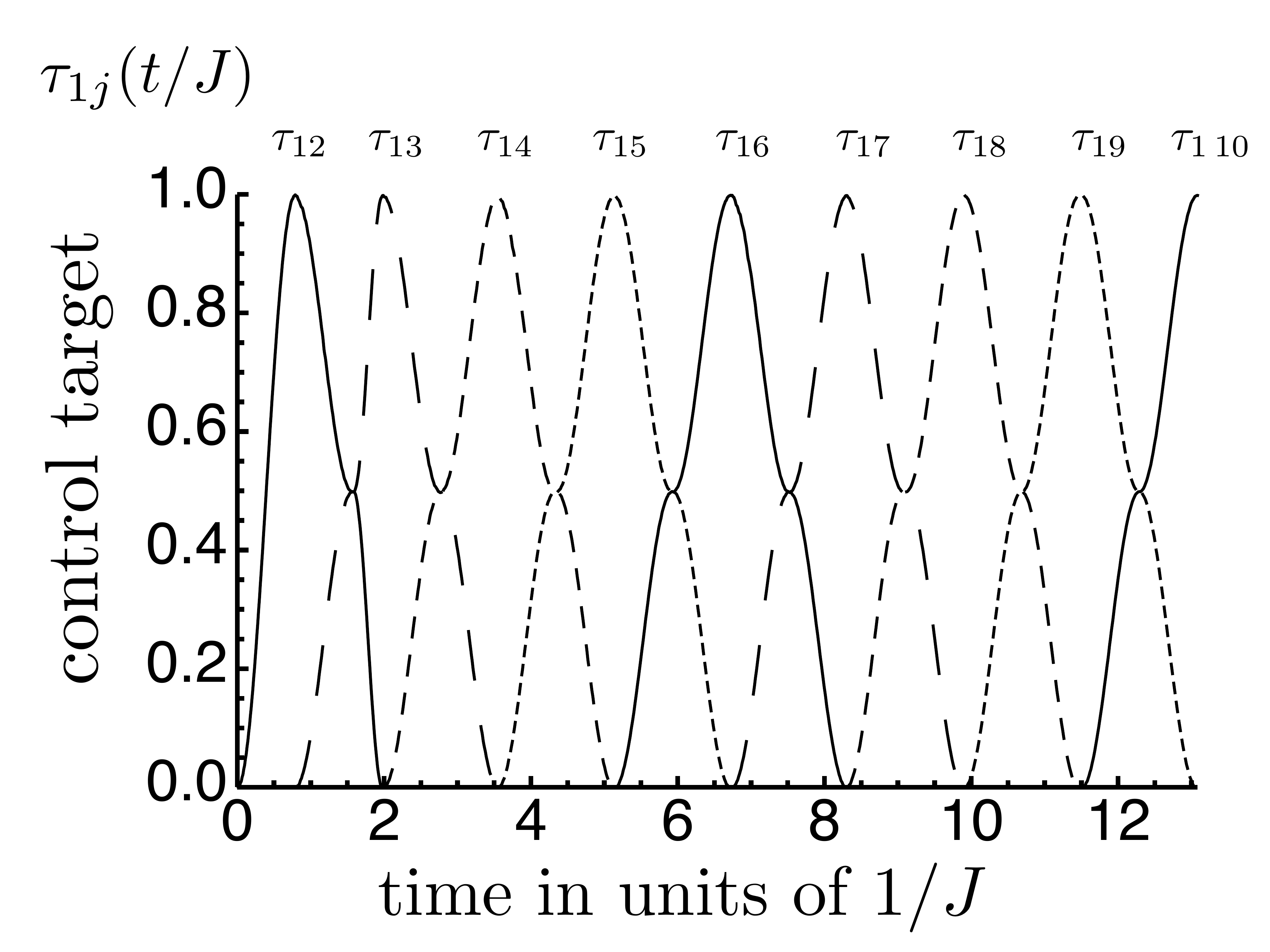}
\caption{Sequence of control targets (equation~\eqref{eq:tau}) as function of time.
$\tau_{12}$ grows until it saturates at $t_1$. At this point $\tau_{12}$ is replaced by $\tau_{13}$ as target functional.
This results in a decrease of $\tau_{12}$ and an increase of $\tau_{13}$, which ends at $t_2$ when, again the target functional is changed.
This process continues until $\tau_{1,10}$ reaches it maximum.}
\label{Fig1}
\end{center}
\end{figure}

Before discussing disordered chains, let us first demonstrate the functionality in terms of an ideal, ordered chain.
Figure \ref{Fig1} depicts the sequential increase and decrease of the different control targets $\tau_{1j}$
with $\mu=0$ and $\alpha_k=1$
for a chain of $N=10$ spins with uniform interactions, {\it i.e.} $J_i=J$ for all $i$.
The control Hamiltonians are limited by $\beta=\sqrt{(g_i^{x})^2+(g_i^{y})^2}=70 J$ and they remain constant over periods of $\Delta=1/(1000J)$
\footnote{There is substantial freedom in the choice for these parameters. Since the time-scales on which entanglement can be generated is limited by the spin-spin-interactions, the performance of control can not be enhanced arbitrarily through larger control amplitudes; it is thus advisable to choose an amplitude that is sufficiently larger than $J$, but sufficiently small so that an integration based on finite time-steps is reliable.
In the choice of a value for $\Delta$ one can choose a compromise between efficiency and accuracy.}.
The chain is initialized in a separable state $|+\rangle^{\otimes N}$, where $|+\rangle$ is the eigenstate of $\sigma^x$ with eigenvalue $+1$,
and the target $\tau_{1j}$ is replaced by $\tau_{1,j+1}$ if $\tau_{1j}$ saturates.

The value of $\tau_{1j}$ is reduced if spins other than $1$ and $j$ participate in any entanglement.
This is merely due to the necessity to keep many-body entanglement sufficiently small for an efficient simulation, but the original goal of creating long-distant entanglement is not jeopardised by an unintentional creation of additional entanglement.
One should therefore characterize the performance by the pairwise entanglement,
{\it i.e.} the entanglement of the reduced density matrix $\varrho_{1j}$ of spins $1$ and $j$, rather than $\tau_{1j}$.
We characterize the pairwise entanglement via Wootters' convex roof construction of concurrence $c_{1j}=\sqrt{\mu_1}-\sum_{i=2}^{4}\sqrt{\mu_i}$ with the decreasingly ordered eigenvalues $\mu_i$ of $(\sigma_y\otimes\sigma_y)\varrho_{1j}^\ast(\sigma_y\otimes\sigma_y)\varrho_{1j}$ \cite{PhysRevLett.80.2245}.

Indeed, one observes a sequential growth and decline of the different selections of pairwise entanglement $c_{1j}$ similar to the behaviour of $\tau_{1j}$ depicted in figure~\ref{Fig1}.
There is an essentially negligible decrease of the peak height as $j$ increases, and substantial entanglement of $c_{1\ 10}\simeq 0.999$ is established between the two spins at the end of the chain. Tests with longer chains gave $c_{1N}\simeq 0.997$ for $N=20$ and $c_{1N}\simeq 0.994$ for $N=40$.
There is thus only a negligible decay of the achievable entanglement with increasing systems size.

\begin{figure}[h]
\begin{center}
\includegraphics[width=0.9\textwidth]{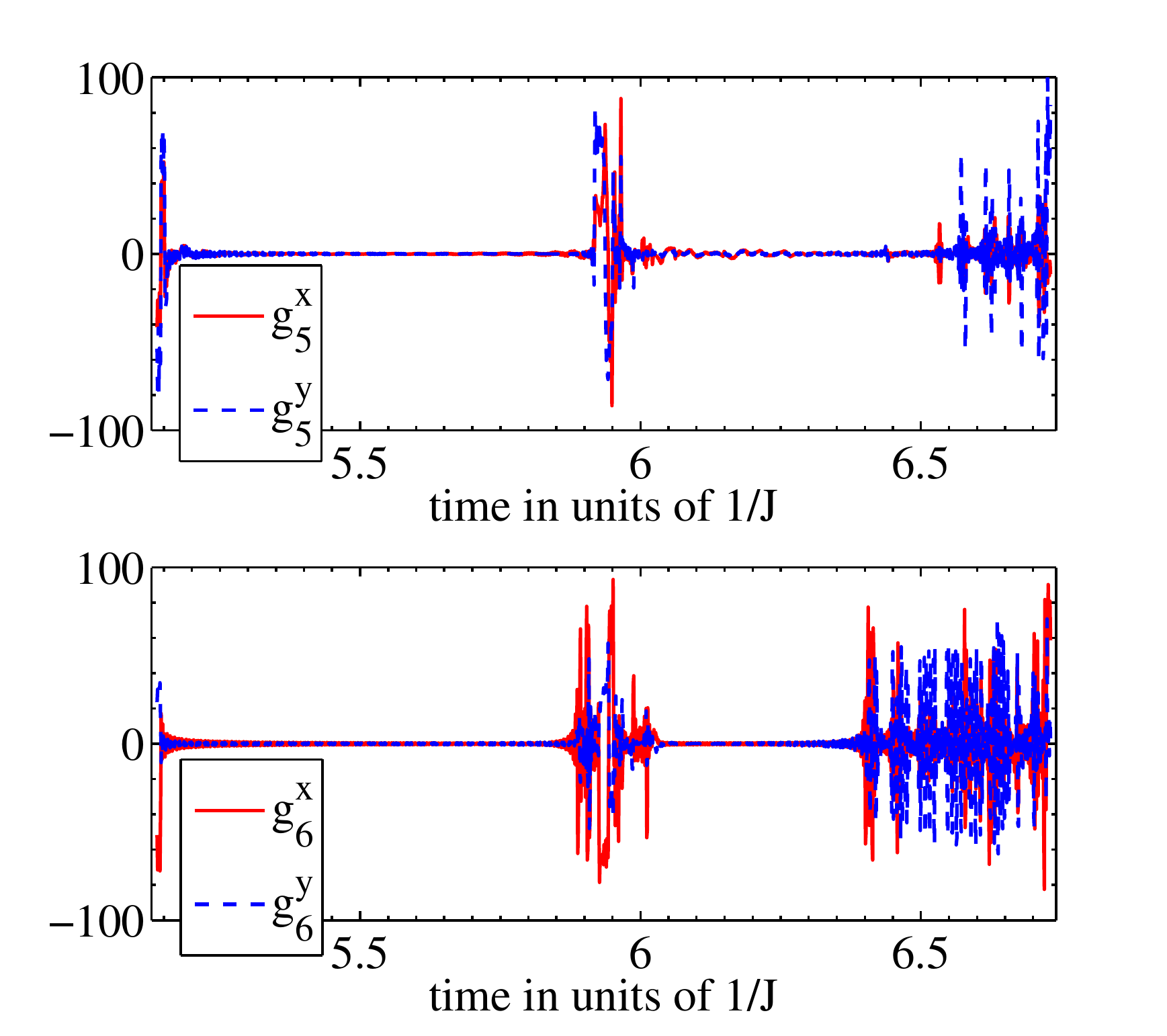}
\caption{Control sequences for spin $5$ and $6$ that realize the swapping from $\tau_{15}$ to $\tau_{16}$ in an ideal Ising chain. The control can be divided into time-intervals in which there is hardly any control and the systems evolves essentially freely and time-intervals in which control is being applied.
This feature is not specific for the homogeneous chain, but we found the same behaviour also for disordered chains, and for the case of reduced control discussed below in section \ref{sec:reducedcontrol}.}
\label{fig_pulse}
\end{center}
\end{figure}

Fig.~\ref{fig_pulse} depicts as an example the time-dependent control (Eq. \ref{eq:optimalparameters}) of spins $5$ to $6$ that achieve the swapping from $\tau_{15}$ to $\tau_{16}$ in an ideal Ising chain.
The construction in terms of finite time steps tends to result in un-necessary high-frequency components;
we explicitly verified that such components can be dropped without sizeable reduction of performance and 
 fig.~\ref{fig_pulse} depicts such a `smoothened' pulse \footnote{The smoothened pulses can also exceed amplitudes of $70J$.}.
 As one can see, the control is mostly limited to short time windows, with extended time windows of free dynamics in-between;
that is, the control tends to align the spins such that their subsequent dynamics exploits the intrinsic interaction in an optimal fashion.

\subsection{Disordered chain}

\begin{figure}[h]
\begin{center}
\includegraphics[width=0.9\textwidth]{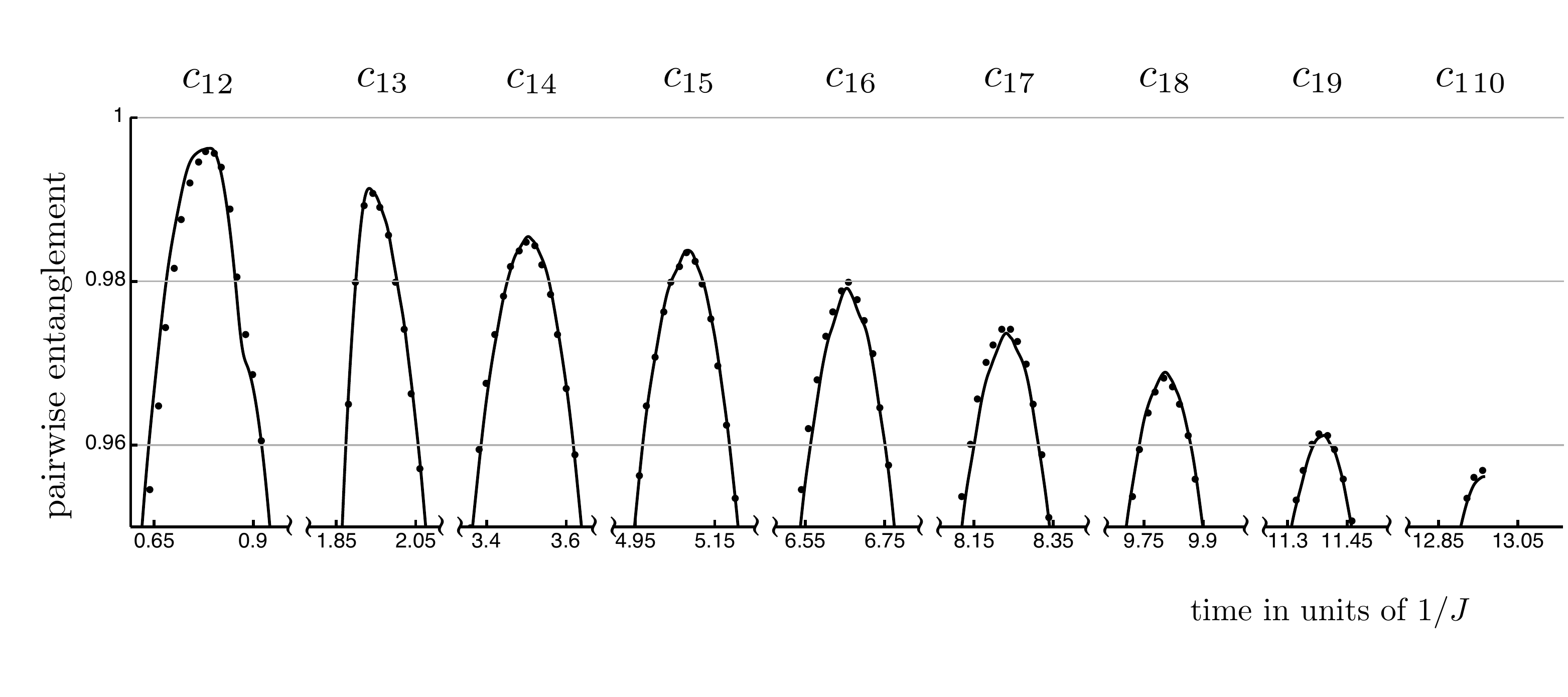}
\caption{Peaks of pairwise entanglement $c_{1j}$
(quantified in term of concurrence \cite{PhysRevLett.80.2245}) that are reached sequentially.
The solid lines depict the average entanglement for an ensemble of $50$ spin chains with disordered coupling constants.
The dots show the average entanglement dynamics for a test-ensemble of $50$ different disordered spin chains resulting from the same control sequence.
There is essentially no drop of pairwise entanglement, {\it i.e.} the performance of the control sequence is largely independent of the specific properties of a spin chain.}
\label{Fig2}
\end{center}
\end{figure}

The performance of the present control methods is by no means specific for uniform chains.
Repeating this procedure for disordered chains with $J_i=r_iJ$ where $r_i$ are random numbers drawn from a uniform distribution $[0.9,1.1]$ resulted in similar values of $c_{1N}$,
and in the following we can address the central question of whether this methods permits to identify a control pulse that works independently of the specific realization of the coupling constants $J_i$.

Since there is typically not a unique optimal (or close to optimal) pulse for a given set of coupling constants,
it is possible to find a pulse that performs well for different interaction landscapes.
We can therefore consider an ensemble of chains with different realizations of coupling constants,
and construct the optimal control Hamiltonians via the ensemble average.
If the utilized ensemble is sufficiently large one can expect the resulting pulse to perform irrespective of the details of the actual system. 
Figure~\ref{Fig2} shows the entanglement dynamics around the peaks of $c_{1j}$ resulting from a control sequence constructed with an ensemble average over $N=50$ different realizations.
Only a minor tribute is paid  to the disorder, as the maximally reached value of $c_{1N}$ is $0.956$; that is, there is a loss of about $4\%$.
To address the question of whether this pulse is applicable to this specific ensemble only, or, if it will perform equally well for any other random realization of coupling constants,
we can apply the pulse to a test ensemble of another $50$ randomly chosen spin chains.
As the dots in figure~\ref{Fig2} show, the behaviour on this test ensemble is hardly different than that of the original ensemble,
what substantiates that the control induces dynamics that is largely insensitive to variations in the interaction landscape.
Despite the fact that our method does not involve iterative refinement of the pulse, it yields good results even for more strongly disordered chains as we explicitly verified for an ensemble with coupling constants drawn from the interval $[0.8,1.2]$;
even though the maximal amplitude of the static noise is comparable in size to the typical interaction constant, one obtains substantial entanglement $c_{1N}\simeq 0.865$ between the two end spins for $N=10$.

\subsection{Reduced control\label{sec:reducedcontrol}}

So far, we considered the control of all spins.
For most of the spins, however, control was introduced only to ensure that MPS are a good description.
We may, therefore also relax the control and choose $\alpha_k=0$ for many spins.

\begin{figure}[h]
\includegraphics[width=0.9\textwidth]{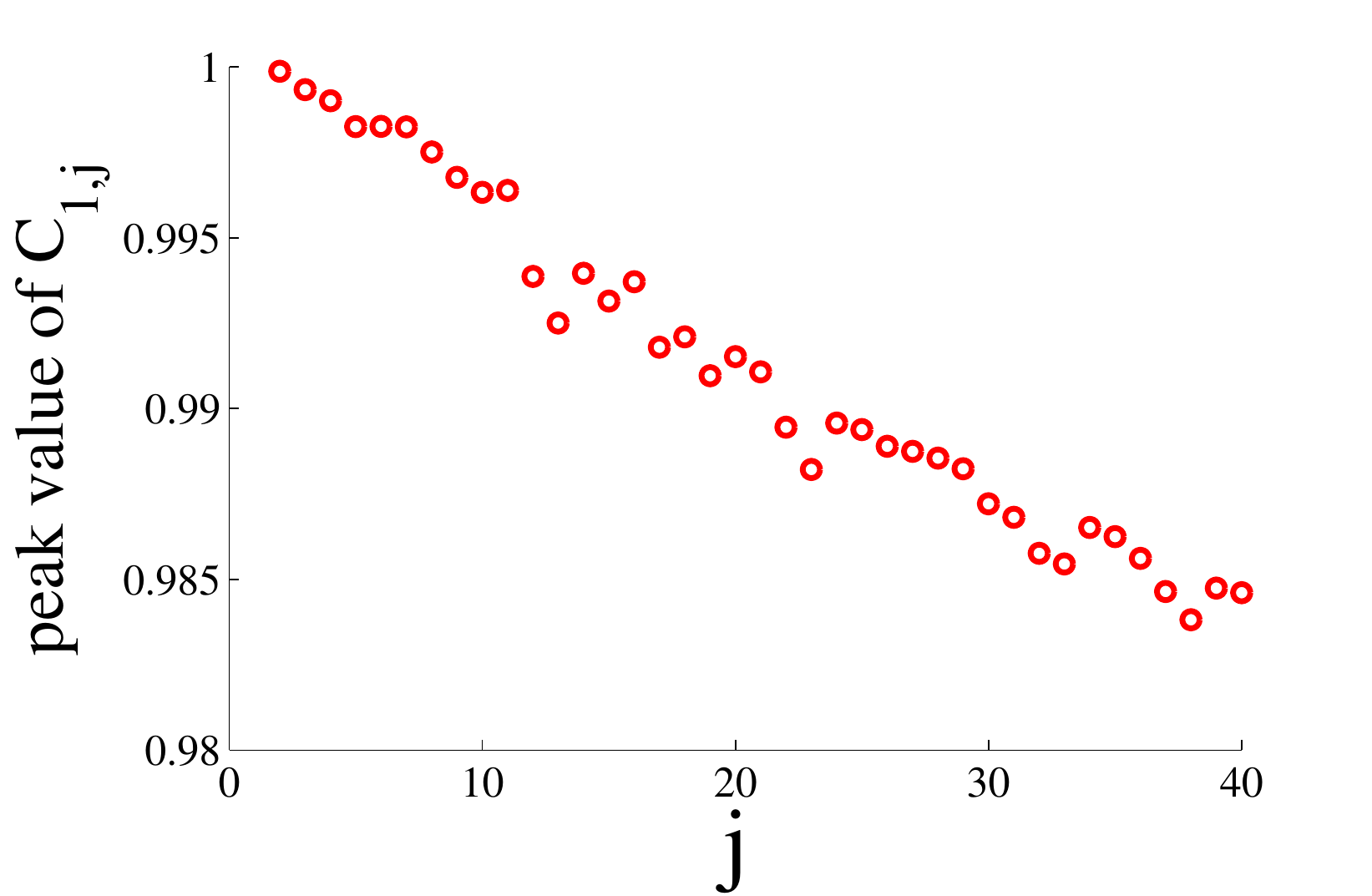}
\caption{Peak values of pairwise entanglement that are achieved sequentially under control only on the $6$ key sites. The controls are obtained by maximizing the second order time derivative of target functional with two-body terms. }
\label{Fig3}
\end{figure}

Quite essential is control of first spin; and during the $j-1^{st}$ interval, in which the entanglement shared with spin $1$ is swapped between the $j-1^{st}$ and the $j^{th}$ spin, also control on spin $j-1$ and $j$ is essential.
It seems plausible that control of spins that are far away from any of these three spins is less important than control of spins that are close by one of the essential spins.
We have investigated the performance of control on a reduced number of spins, and found that control of many spins can be forfeited.
Quite surprisingly, control of spin $2$ is not necessary for good performance after the third interval.
Control of $j-2^{nd}$ and $j+1^{st}$ however is important during the swapping procedure form spin $j-1$ to spin $j$.
Spins that had participated in a swapping operation will not become relevant any more; it is therefore not surprising that control on spins $<j-2$ can readily be given up.
However, spins that will participate in a swapping operation in the near future need to be controlled and we found that control on spin $j+2$ is required for good performance.
One may thus reduce the control to spin $1$ and spins $j-2$ through $j+2$ with $j$ increasing by $1$ after each swapping operation.
With control on these $6$ spins only, one obtains very good performance as depicted in Fig.\ref{Fig3}.
The loss of entanglement during the swapping operations is essentially negligible and substantial entanglement $c_{1\ 40}\simeq 0.985$ is established over a chain of $40$ spins.

Given the control on a reduced number of spins, a thorough check of the accuracy of the numerical propagation is in order.
We have therefore simulated the dynamics with MPS of different bond dimension.
The entanglement that builds up during the dynamics suggests a necessary minimal bond dimension of $8$.
We worked with a bond dimension of $10$, and explicitly confirmed that an increase to a bond dimension of $20$ does not result in any discernible change in dynamics, what confirms the validity of the MPS description with low bond dimension.

\section{Discussions}
It is interesting to notice that our approach is quite different from the notion of `perfect state transfer' (PST) or `almost perfect state transfer' (APST), where two parties employ a spin chain with perfect coupling strength as quantum wire.
The protocol is initialized with the preparation of the first spin in the to-be-transferred state.
After some period of evolution induced by the system Hamiltonian, the final spin will have this state with non-zero fidelity \cite{Bose_PST,PhysRevA.84.012307,PhysRevLett.109.050502}.
This protocol certainly also permits to create some distant entanglement, but our control scheme has the advantages that:
i) It is robust against disorder in the coupling of the spins and against the lengths of the spin chain, whereas (A)PST greatly depends on the coupling and the length of the spin wire \cite{PhysRevA.84.012307,PhysRevLett.109.050502}.
ii) The amount of entanglement that can be achieved with the present control scheme is substantially higher than that of (A)PST, and
iii) the time cost to achieve such high entanglement is much lower.
For example, the maximally achieved entanglement between site $1$ and site $10$ within time cost $4000/J$ in a perfect spin chain by means of (A)PST is $0.95$ \cite{Bose_PST} as compared to
$0.999$ for a perfect spin chain and $0.958$ for a disordered spin chain that can be created within $12.76/J$ with control.
In particular with increasing system size, the advantage of the present method becomes apparent:
entanglement established over $80$ sites within time cost $4000/J$ in a perfect spin chain with (A)PST does not exceed $0.5$ \cite{Bose_PST}, whereas using our control strategy permits to create entanglement between site 1 and site 80 amounting to $0.990$ with the time cost roughly equal to $130/J$.

The creation of long-distance entanglement is also by no means limited to bipartite entanglement,
but may also be employed for the creation of entanglement between three or more distant spins.
The demonstration of the usability of MPS for the control of large spin systems, in particular, also suggests that other goals like the implementation of multi-qubit quantum gates involving distant spins can be realized in a similar fashion.
In all such situations, the applicability of MPS for a given situation can be ensured through a suitably extended target functional that makes sure that many-body entanglement remains sufficiently low.
As simple control strategies like Lyapunov control might fail to identify goals whose realization requires many elementary interactions,
it helps to define a sequence of intermediate goals.
In the present case we did so by sudden changes of the target functional,
but also smooth, continuous modulations of targets (which itself might become object of optimization) is conceivable.
Such well-designed dynamical goals together with advanced numerical techniques like MPS promise to help us make the step from small scale proof-of-principle demonstrations towards large-scales.

\ack

The numerical work was strongly supported by
Mari Carmen Ba\~nuls. J.C acknowledges the hospitality by the Institute for Interdisciplinary Information Sciences Tsinghua University and the Institute of Physics, Chinese Academy of Sciences.
Computational resources in {\em Rechenzentrum Garching}
and financial support by the {\em European Research Council} within the project `Optimal dynamical control of quantum entanglement' is gratefully acknowledged.

\clearpage
\bibliographystyle{unsrt}

\bibliography{reference}

\end{document}